\documentclass[prd,aps,showpacs,preprintnumbers,amssymb]{revtex4}

\usepackage{color}
\usepackage{epsf}

\def\e3p{$\eta \rightarrow 3 \pi$}

\begin{document}
\title{%
\hfill{\normalsize\vbox{%
\hbox{}
 }}\\
{Vacuum energy in some particular quantum field theories}}

\author{Renata Jora
$^{\it \bf a}$~\footnote[2]{Email:
 rjora@theory.nipne.ro}}

\affiliation{$^{\bf \it a}$ National Institute of Physics and Nuclear Engineering PO Box MG-6, Bucharest-Magurele, Romania}

\date{\today}

\begin{abstract}
We study the properties of a class of quantum field theories endowed with an equal number of anti commuting and commuting field variables, the most common example being the supersymmetric models.
Based on the scaling properties of the partition function we find that these theories have the quantum energy momentum tensor associated to dilatations and thus the vacuum energy  equal to zero.

\end{abstract}
\maketitle

\section{Introduction}
The vacuum energy of a quantum field theory has been over the time the subject of intensive debates \cite{Jaffe1}-\cite{Sola} in special in connection to the cosmological constant problem \cite{Carroll}-\cite{Dyson}. Since in the  Minkowski space QFT one can always add a constant to the energy density of the vacuum without spoiling in any way the consequences of the theory it seems superfluous to even try to compute it. However a QFT theory couples to gravity case in which the exact definition of the vacuum is important. In considering the vacuum one can make different assumptions with two main emerging directions: a) one is to consider all quantum fluctuations associated to a Lagrangian including thus the kinetic contributions; b) the other would be to consider only the contributions coming from the scalar potential of the theory. Here we shall adopt the first point of view. We thus define:
\begin{eqnarray}
TE=\frac{\int d \Phi [\int d^4 x{\cal H}]\exp[i\int d^4 x{\cal L}]}{\int d\Phi \exp[i\int d^4 x{\cal L}]}.
\label{deft5665}
\end{eqnarray}
Here $d\Phi$ is a generic notation for all the variable of integration that appear in the Lagrangian ${\cal L}$. Moreover $T$ is the interval of time, ${\cal H}$ is the Hamiltonian density and $E$ is the energy of the vacuum. From the point of view of a quantum field theory the definition in Eq. (\ref{deft5665}) is the safest and the most sensible choice one can make to take into account all quantum fluctuations associated to a specific Lagrangian.

By only looking at Eq. (\ref{deft5665}) one can estimate that the vacuum energy or density ($\rho_V$) is very large. The most common ballpark figure will be that $\rho_V\approx\Lambda^4$ where $\Lambda$ is the cut off associated to a particular theory. It turns out that for a typical theory like the pure standard model the  calculations lead in great measure to this exact estimate. Here however we shall consider a particular case of quantum field theories for which the number of fermion degrees of freedom matches exactly the number of scalar degrees of freedom such that the partition function (without additional factors in the measure of integration) is a simple constant that does not depend on any dimensionful quantity. Note that this equality should function off-shell and includes automatically auxiliary fields. The theories we are referring may be non supersymmetric or supersymmetric theories where the supersymmetry  may be unbroken or broken however badly.  The independence of the partition function on the momenta will be heavily used in what follows. Moreover the matching of degrees of freedom in the partition function suggests that this has some scaling properties.  Relying on these two features of the partition function  we will show in section II that the the energy associated to the vacuum according to Eq. (\ref{deft5665}) and hence the vacuum energy density are zero for this particular class of theories. Section III is dedicated to Discussion.

\section{Vacuum energy of a particular class of quantum field theories}

We consider a typical scalar theory with the Lagrangian:
\begin{eqnarray}
{\cal L}=\frac{1}{2}\partial^{\mu}\Phi\partial_{\mu}\Phi-\frac{1}{2}\Phi^2-V(\Phi),
\label{scalart45}
\end{eqnarray}
where $V(\Phi)$ is the potential.
First we shall review a few known facts regarding the behavior under translations.  A constant translation of the type:
\begin{eqnarray}
x_{\mu}'=x_{\mu}-a_{\mu}
\label{res5546}
\end{eqnarray}
will lead to a transformation of the field  $\Phi$ of the form:
\begin{eqnarray}
\Phi(x')'=\Phi(x)=\Phi(x+a)=\Phi(x')+a^{\mu}\partial_{\mu}\Phi(x')
\label{reee98}
\end{eqnarray}
Under the translations the Lagrangian density will behave as:
\begin{eqnarray}
{\cal L}(x)={\cal L}(x'+a)={\cal L}(x)+a^{\nu}\partial_{\mu}[g^{\mu}_{\nu}{\cal L}].
\label{tr5546}
\end{eqnarray}
Alternatively we can compute the variation of the Lagrangian stemming from the fields:
\begin{eqnarray}
&&\delta {\cal L}=\frac{\delta {\cal L}}{\delta \Phi}\delta\Phi+\frac{\delta{\cal L}}{\delta(\partial_{\mu}\Phi}\delta(\partial_{\mu}\Phi)=
\frac{\delta {\cal L}}{\delta \Phi}a^{\mu}\partial_{\mu}\Phi+\frac{\delta{\cal L}}{\delta(\partial_{\mu}\Phi)}a^{\nu}\partial_{\mu}\partial_{\nu}\Phi=
\nonumber\\
&&[\frac{\delta {\cal L}}{\delta \Phi}-\partial_{\mu}(\frac{\delta {\cal L}}{\delta \partial_{\mu}\Phi})]a^{\nu}\partial_{\nu}\Phi+
\partial_{\mu}[\frac{\delta {\cal L}}{\delta \partial_{\mu}\Phi}\partial_{\nu}\Phi]a^{\nu},
\label{trs4353}
\end{eqnarray}
where we integrated by parts.
From Eqs. (\ref{tr5546}) and (\ref{trs4353}) we deduce that if the equations of motion are satisfied the quantity conserved is the energy momentum tensor:
\begin{eqnarray}
T^{\mu}_{\nu}=\frac{\delta {\cal L}}{\partial_{\mu}\Phi}\partial_{\nu}\Phi-{\cal L}g^{\mu}_{\nu}
\label{ener5466}
\end{eqnarray}
such that $\partial_{\mu}T^{\mu}_{\nu}=0$.

Now we shall extend this to the more general case of space time dependent translations. Then the variations of the coordinate and of the fields are given by:
\begin{eqnarray}
&&x_{\mu}=x_{\mu}'+a_{\mu}
\nonumber\\
&&\Phi(x)=\Phi(x')+a^{\mu}\partial_{\mu}\Phi
\nonumber\\
&&\partial_{\nu}\Phi(x)=\partial_{\nu}\Phi(x')+a^{\mu}\partial_{\mu}\partial_{\nu}\Phi+\partial_{\nu}a^{\mu}\partial_{\mu}\Phi(x')
\label{genre4553}
\end{eqnarray}
Note that the third term on the right hand side of the third equation in (\ref{genre4553}) was absent in the case of constant translations.  This term will contribute additionally to the variation of the Lagrangian as it cannot be included in $a^{\nu}\partial_{\mu}[g^{\mu}_{\nu}{\cal L}]$. Thus we will have on one hand:
\begin{eqnarray}
\delta {\cal L}=a^{\nu}\partial_{\mu}[g^{\mu}_{\nu}{\cal L}]+\partial_{\mu}a^{\nu}\partial^{\mu}\Phi\partial_{\nu}\Phi,
\label{var12131}
\end{eqnarray}
and on the other,
\begin{eqnarray}
&&\delta {\cal L}=\frac{\delta {\cal L}}{\delta \Phi}\delta\Phi+\frac{\delta{\cal L}}{\delta(\partial_{\mu}\Phi)}\delta(\partial_{\mu}\Phi)=
\nonumber\\
&&\frac{\delta {\cal L}}a^{\mu}\partial_{\mu}\Phi+\frac{\delta{\cal L}}{\delta(\partial_{\mu}\Phi)}(a^{\nu}\partial_{\mu}\partial_{\nu}\Phi+
\partial^{\nu}a_{\mu}\partial^{\mu}\Phi\partial_{\nu}\Phi)=
\nonumber\\
&&[\frac{\partial {\cal L}}{\partial \Phi}-\partial_{\mu}(\frac{\delta {\cal L}}{\delta \partial_{\mu}\Phi})]a^{\nu}\partial_{\nu}\Phi+
\partial_{\mu}[\frac{\delta{\cal L}}{\delta (\partial_{\mu}\Phi)}a^{\nu}\partial_{\nu}\Phi].
\label{res5536}
\end{eqnarray}
Since  for a space time dependent translations we consider the variation of the action and not that of the Lagrangian the last term on the right hand side of the last equation in Eq. (\ref{res5536}) will vanish because is a total derivative. Then by equating Eqs. (\ref{var12131}) with (\ref{res5536}) we obtain:
\begin{eqnarray}
&&\delta S=\int d^4 x'[a^{\nu}\partial_{\mu}[g^{\mu}_{\nu}{\cal L}]+\partial^{\nu}a_{\mu}\partial^{\mu}\Phi\partial_{\nu}\Phi]=
\nonumber\\
&&=\int d^4 x[\frac{\partial {\cal L}}{\partial \Phi}-\partial_{\mu}(\frac{\delta {\cal L}}{\delta \partial_{\mu}\Phi})]a^{\nu}\partial_{\nu}\Phi.
\label{res44235}
\end{eqnarray}
Furthermore we can write the first line in the above equation as:
\begin{eqnarray}
\delta S=\int d^4 x'[-\partial_{\mu}a^{\nu}[g^{\mu}_{\nu}{\cal L}]+\partial_{\mu}a^{\nu}\partial^{\mu}\Phi\partial_{\nu}\Phi]= \int d^4 x' (\partial_{\mu}a^{\nu})T^{\mu}_{\nu}.
\label{res1231}
\end{eqnarray}
If the equation of motion are satisfied one obtains the conservation of the energy momentum tensor as before.

The findings obtained so far extend to all possible fields and interaction in a general Lagrangian. In what follows we shall consider one particular type of Lagrangians in which the number of bosonic dgerees of freedom is equal to that of fermion ones. This can be a supersymmetric Lagrangian or one like that introduced in \cite{Jora}. We shall start by writing the partition function for this Lagrangian:
\begin{eqnarray}
Z=\int [d\Phi](x)[d A](x) [d\Psi](x) [dc](x)\exp[i d^4x {\cal L}(\Phi(x),A(x),\Psi(x),c(x))]
\label{part3456}
\end{eqnarray}
where by $[d\Phi]$, $[d A]$, $[d \Psi]$, $[dc]$ we denoted generically all scalar fields, gauge fields, fermion fields or ghost that might appear in such a Lagrangian. The partition function $Z$ can be expresses also as:
\begin{eqnarray}
Z=\int[d\Phi](x)[d A](x) [d\Psi](x) [dc](x)\exp[i d^4x' {\cal L}(\Phi(x'),A(x'),\Psi(x'),c(x'))+ \int d^4 x' \partial_{\mu}a^{\nu}T^{\mu}_{\nu}(x')].
\label{es4355252}
\end{eqnarray}
Note that the measure of integration is expressed in terms of the of the coordinate $x'$. One might argue that the measure of integration is independent of the translations of the coordinate. This would require additional factors in the measure and delicate mathematical arguments and derivation. Here we shall adopt a more mundane point of view suggesting that the transformation from $[d \Phi(x)]$ (or all the other factors that appear in the measure of integration) is done as for  regular variables  and requires the computation of simple Jacobians. In the Appendix A  we show that for the case when the number of boson degrees of freedom is equal to that of fermions these Jacobians cancel and that indeed:
\begin{eqnarray}
[d\Phi](x)[d A](x) [d\Psi](x) [dc](x)=[d\Phi](x')[d A](x') [d\Psi](x') [dc](x').
\label{res5546}
\end{eqnarray}
This implies:
\begin{eqnarray}
&&Z=\int [d\Phi](x')[d A](x') [d\Psi](x') [dc](x')\exp[i d^4x' {\cal L}(\Phi(x'),A(x'),\Psi(x'),c(x'))+ \int d^4 x' (\partial_{\mu}a^{\nu})T^{\mu}_{\nu}(x')]=
\nonumber\\
&&\int [d\Phi](x)[d A](x) [d\Psi](x) [dc](x)\exp[i d^4x {\cal L}(\Phi(x),A(x),\Psi(x),c(x))].
\label{res664774}
\end{eqnarray}
From now on we shall drop the superscript in $x'$ and  consider $x'=x$. First we write the measure of integration and the action on a lattice in the Fourier space  for which we exemplify only two terms:
\begin{eqnarray}
&&\int d^4x \frac{1}{2}\partial^{\mu}\Phi\partial_{\mu}\Phi=\frac{1}{2V}\sum_n\Phi(-k_n)\Phi(k_n)k_n^2
\nonumber\\
&&\int d^4x (\partial_{\mu}a^{\nu})T^{\mu}_{\nu}(x)=\frac{i}{V}\sum_m ik_{m\mu}a^{\nu}(-k_m)T^{\mu}_{\nu}(k_m),
\label{resa2121}
\end{eqnarray}
where $V$ is the space time volume.
Then we  notice that the partition function is independent of the arbitrary real small parameters $a^{\nu}(-k_m)$ (we can always arrange if $a^{\nu}(x)$ is small to have also its Fourier transform small):
\begin{eqnarray}
&&\frac{\delta Z}{\delta a^{\nu}(-k_m)}|_{a^{\nu}(-k_m)=0}=0=
\nonumber\\
&&\prod_k [d\Phi](k_k)[d A](k_k) [d\Psi](k_k) [dc](k_k)[\frac{i}{V}\sum_m ik_{m \mu}(-k_m)T^{\mu}_{\nu}(k_m)]\exp[i d^4x {\cal L}(\Phi(x),A(x),\Psi(x),c(x))].
\label{res664554}
\end{eqnarray}
which shows that $\langle \Omega|\partial_{\mu}T^{\mu}_{\nu}|\Omega\rangle=0$ even if we do not require for the equations of motion to be satisfied.

Next since the parameter $a_{\mu}(x)$ is small and arbitrary we can pick it however we want and we shall consider it of the form  $a_{\mu}(x)=\epsilon x_{\mu}$ where $\epsilon$ is a very small parameter. Here we assume that the space time volume although very large is finite and we denote it by $L^4$ such that we can always pick an $\epsilon$ such that $\epsilon L$ is still very small. Then,
\begin{eqnarray}
\int d^4 x (\partial_{\mu}a^{\nu})T^{\mu}_{\nu}(x)=\int d^4 x(-\epsilon) g^{\nu}_{\mu}T^{\mu}_{\nu}.
\label{res442321}
\end{eqnarray}
Since the partition function is again independent of the small parameter $\epsilon$ we further write:
\begin{eqnarray}
\frac{\delta Z}{\delta \epsilon}|_{\epsilon=0}=\int [d\Phi](x)[d A](x) [d\Psi](x) [dc](x)[\int  d^4 x g^{\nu}_{\mu}T^{\mu}_{\nu}]\exp[i d^4x {\cal L}(\Phi(x),A(x),\Psi(x),c(x))]=0,
\label{res423324}
\end{eqnarray}
to deduce that $\langle \Omega |\int d^4 xg^{\nu}_{\mu}T^{\mu}_{\nu}|\Omega\rangle=0$.

Next we switch back to the Fourier space for the partition function to notice that in the theory with matching number of boson and fermion degrees of freedom (without additional factors in the measure of integration) the partition function is completely independent of the individual momenta. Then one can write:
\begin{eqnarray}
\frac{\delta Z}{\delta k_{\nu p}}=\prod_k [d\Phi](k_k)[d A](k_k) [d\Psi](k_k) [dc](k_k)[i\Phi(-k_p)k_{p\nu}\Phi(k_p)+....]\exp[iS]
\label{res55544}
\end{eqnarray}
The dots stand for additional kinetic terms of all the particles involved. We can further multiply Eq.(\ref{res55544}) by $k_{p}^{\mu}$ and sum over all the momenta to get:
\begin{eqnarray}
0=\prod_k [d\Phi](k_k)[d A](k_k) [d\Psi](k_k) [dc](k_k)[i\sum_p\Phi(-k_p)k_{p\nu}k_p^{\mu}\Phi(k_p)+....]\exp[iS]
\label{res4435667}
\end{eqnarray}
to get that (expressed again in the coordinate space): $\langle\Omega |\int d^4 x\frac{\delta {\cal L}}{\delta \partial_{\mu}\Phi}\partial_{\nu}\Phi+.....|\Omega\rangle=0$. But adding the terms corresponding to all the fields we obtain that that part of the energy momentum tensor that contains derivative with respect to the derivatives of the fields is zero.

In summary  from Eqs. (\ref{res664554}), (\ref{res423324}) and (\ref{res4435667}) we derive the following results:
\begin{eqnarray}
&&\langle \Omega|\partial_{\mu}T^{\mu}_{\nu}|\Omega\rangle=0
\nonumber\\
&&\langle \Omega |\int d^4 x g^{\nu}_{\mu}T^{\mu}_{\nu}|\Omega \rangle=0
\nonumber\\
&&\langle \Omega|\int d^4 x (T^{\mu}_{\nu})_1|\Omega\rangle=0
\label{results5546}
\end{eqnarray}
where
\begin{eqnarray}
(T^{\mu}_{\nu})_1=\frac{\delta {\cal L}}{\partial_{\mu}\Phi}\partial_{\nu}\Phi+{\rm similar\,\,terms\,\,for\,\,all\,\,the\,\,fields\,\,involved}.
\label{somre435}
\end{eqnarray}

In particular from the second and third lines (where in third line we pick $\mu=\nu$)  in Eq. (\ref{results5546}) one obtains:
\begin{eqnarray}
&&\langle \Omega|\int d^4 x T^{\mu}_{\nu}|\Omega\rangle=0
\nonumber\\
&&\langle \Omega |\int d^4 x {\cal L}|\Omega\rangle=0
\nonumber\\
&&\langle\Omega|\int d^4 x {\cal H}|\Omega\rangle=0,
\label{res664756}
\end{eqnarray}
since the third line of Eq. (\ref{results5546}) is true for each combination of the space time indices and we know that:
\begin{eqnarray}
{\cal H}=T^{00}.
\label{res54646}
\end{eqnarray}
Note that this this Hamiltonian includes contribution from the auxiliary fields in the theory (see Appendix B for an example).
\section{Discussion}

In this paper we showed that the vacuum energy density in a theory with matching commuting and anticommuting degrees of freedom is equal to zero. This may or may not have bearing on issues related to the smallness of the cosmological constant $\rho_{\Lambda}=1.48\pm0.11\times 10^{-123}\rho_{Planck}$ (\cite{Riess}, \cite{Perlmutter}, \cite{Tegmark}) for several reasons. First we could not in our approach generate any small quantity. If one considers a model with mismatching degrees of freedom  then the corresponding vacuum energy density will be proportional to $\Lambda^4f(N)$ where $\Lambda$ is the cut-off of the theory and $f(N)$ is an integer function of the degrees of freedom  and thus cannot be small.  Thus our results agree with the standard estimates for the vacuum energy density for theories outside the class considered here.  On the other hand our results refer  exclusively to the Minkowski space and we do not know in what measure a curved space time may alter them. According to some authors \cite{Sola} it is possible that the space time curvature may have tiny effects on the vacuum energy. If this is the case in our context the smallness of the cosmological constant may be associated to these effects.

Finally as an aside to the standard model one should discuss the contribution of various vacuum condensates that might appear in the theory. These are not included in our approach. We rely on the idea that QCD vacuum condensates are an intrinsic property of the hadrons and should not appear as additional contribution to the vacuum energy. Electroweak theory comes with its own vacuum but in order for this to compensate for  the vacuum energy contribution coming from the mismatch of degrees of freedom it should be of order $\rho_{EW}\approx(200 \,Gev)^4\approx \Lambda^4$ where $\Lambda$ is the associated cut-off of the theory. For such a delicate cancellation to occur such that $\rho_{\Lambda}$ has the small values that we know today one would need in the theory a cut-off of the order $\Lambda \approx 200\, GeV$ which may be realistic or not.  However the electroweak symmetry breaking may still have an underlining strong dynamics, as this situation has not been completely excluded by the LHC, case in which $\rho_{EW}$ remains unknown or it is zero.

In conclusion we found that the vacuum energy of a theory with matching  commuting and anticommuting degrees of freedom is equal to zero with the provision that this energy includes the contribution coming from the auxiliary states.  For a theory with mismatching degrees of freedom the vacuum energy density is proportional to $\Lambda^4f(N)$ where $\Lambda$ is the cut-off of the theory and $f(N)$ is an integer linear function of the degree of freedom. These results may have significance for estimating QFT contributions to $\rho_{\Lambda}$ or from a pure theoretical point of view.

\section*{Acknowledgments} \vskip -.5cm

The work of R. J. was supported by a grant of the Ministry of National Education, CNCS-UEFISCDI, project number PN-II-ID-PCE-2012-4-0078.

\begin{appendix}

\section{}
Consider the change due to translations given in Eq. (\ref{genre4553}):
\begin{eqnarray}
\Phi(x)=\Phi(x')+a^{\mu}\partial_{\mu}\Phi(x').
\label{ch66578}
\end{eqnarray}
We shall now write the right hand side of Eq. (\ref{ch66578}) in the Fourier space on a lattice:
\begin{eqnarray}
&&\Phi(x')+a^{\mu}\partial_{\mu}\Phi(x')=
\nonumber\\
&&\frac{1}{V}\sum_n \Phi(k_n)\exp[-ik_nx']+\frac{1}{V^2}\sum_p a^{\mu}(k_p)\exp[-ik_px']\partial_{\mu}[\sum_t\Phi(k_r)\exp[-ik_rx']=
\nonumber\\
&&\frac{1}{V}\sum_n[\phi(k_n)+\sum_ra^{\mu}(k_n-k_r)(-ik_r)\Phi(k_r)]\exp[-ik_nx'],
\label{res44355}
\end{eqnarray}
where we renamed $k_p=k_n-k_r$. Then the Jacobian in the Fourier space associated to the real scalar field $\Phi$ is:
\begin{eqnarray}
&&J_{\Phi}=\big|\frac{\delta \Phi(k_n)}{\delta \Phi (k_p)}\big |=
\nonumber\\
&&\big |\delta_{np}+\frac{1}{V} a^{\mu}(k_n-k_p)(-ik_{p\mu}) \big |.
\label{jac44355}
\end{eqnarray}
The main point here is that the jacobian does not depend on the field $\Phi$ so it is universal.

Note that under the same translations the fermions field transforms as:
\begin{eqnarray}
\Psi(x)=\Psi(x') + a^{\mu}\partial_{\mu}\Psi(x')
\label{fer443566}
\end{eqnarray}
so the fermion fields will induce the same type of Jacobian at an inverse power.

The gauge fields transform as:
\begin{eqnarray}
&&A_{\nu}(x)=\frac{\partial x^{\prime \mu}}{\partial x^{\nu}}A_{\mu}(x')=g^{\mu}_{\nu}+(\partial^{\mu}a_{\nu})A_{\mu}.
\label{res554667}
\end{eqnarray}
 Eq.(\ref{res554667}) can be expressed in the Fourier space as before and leads to the Jacobian:
\begin{eqnarray}
&&J_{A_{\mu}}=\big |\frac{\delta A^{\rho}(k_n)}{\delta A^{\nu} (k_p)}\big |=
\nonumber\\
&&\det[g^{\rho}_{\nu}\delta_{np}+\frac{1}{V}a^{\rho}(k_n-k_p)(-ik_{p\mu}+ik_{n\mu})g^{\mu}_{\nu})]=
\nonumber\\
&&\det[g^{\rho}_{\nu}]\det[\delta_{np}+\frac{1}{V}a^{\nu}(k_n-k_p)(-ik_{p\nu}+ik_{n\mu})].
\label{res443566}
\end{eqnarray}
We are interested in linear contribution in $a_{\nu}$ from these determinants. These automatically come up with a trace over space time indices and the momenta (which is equivalent to an integral over the momenta).  But in this case the only possible contributions that survive are those of the type $a^{\nu}(k_n-k_p)(-ik_{p\nu}+ik_{n\mu})$ where $a^{\nu}$ and the factor in front depend on the same momenta. This is true because an integral over an odd number of momenta is otherwise zero. Note that this includes also the situation when the Fourier transform of $a^{\mu}$ may not be a proper function but rather a functional. We conveniently extract these combinations from all Jacobians to note that the number of positive contributions coming form the boson variables cancels exactly the contributions coming from anticommuting ones. This is due to to the exact match of fermion (anticommuting) and boson degrees of freedom.


\section{}

Consider the simplest supersymmetric model, a free chiral supermultiplet. We add to the corresponding Lagrangian a mass term for the scalars that breaks supersymmetry to obtain the Lagrangian:

\begin{eqnarray}
{\cal L}=-\partial^{\mu}\Phi^*\partial_{\mu}\Phi+i\Psi^{\dagger}\bar{\sigma}^{\mu}\Psi+m^2\Phi\Phi^*+FF^*.
\label{lres4435}
\end{eqnarray}
We will show first that indeed according to section II the following result holds:
\begin{eqnarray}
\langle \Omega|\int d^4 x {\cal L}|\Omega\rangle=0.
\label{res221312}
\end{eqnarray}

For that we introduce the dummy parameters $k_1$, $k_2$, $k_3$ in the Lagrangian such that this will become:
\begin{eqnarray}
{\cal L}_1=-k_1\partial^{\mu}\Phi^*\partial_{\mu}\Phi+ik_2\Psi^{\dagger}\bar{\sigma}^{\mu}\Psi+m^2\Phi\Phi^*+k_3FF^*.
\label{res5557865}
\end{eqnarray}
The partition function for this Lagrangian can be computed easily by going to the Fourier space and computing the gaussian integrals:
\begin{eqnarray}
&&Z=\int d \Phi d \Phi^* d\Psi d \Psi^{\dagger}d F dF^* \exp[i\int d^4 x {\cal L}]\approx
\prod_{p_{n\mu}}[\frac{1}{-k_1p_n^2+m^2}{k_2^2p_n^2}\frac{1}{k_3}].
\label{part45546}
\end{eqnarray}
Then:
\begin{eqnarray}
&&\langle \Omega|\int d^4 x {\cal L}|\Omega\rangle=
\nonumber\\
&&\langle \Omega|\int d^4 x {\cal L_1}|\Omega\rangle_{k_1=k_2=k_3=1}=
\nonumber\\
&&\big[k_1\frac{d \ln Z}{d k_1}+k_2\frac{d \ln Z}{d k_2}+k_3\frac{d \ln Z}{d k_3}+m^2\frac{ d\ln Z}{ d m^2}]|_{k_1=k_2=k_3=1}=
\nonumber\\
&&[\sum_{n}(-\frac{1}{-k_1p^2+m^2})(-k_1p^2)+\sum_n (2)+\sum_n(-1)+\sum_n(-\frac{1}{-k_1p^2+m^2})m^2]|_{k_1=k_2=k_3=1}=
\nonumber\\
&&\sum_n[(-1)+\sum_n(2)+\sum_n(-1)]|_{k_1=k_2=k_3=1}=0.
\label{proof65775}
\end{eqnarray}
Here $\sum_n$ is the sum over all lattice points in a Fourier space for momenta and $\sum_n=V\int d^4\frac{1}{(2\pi)^4}$ where $V$ is the space time volume. We also used the simple relation:
\begin{eqnarray}
x\frac{d\prod_i a_i}{dx}=\prod_i a_i \sum_i \frac{x}{a_i}\frac{d a_i}{d x}.
\label{us44356}
\end{eqnarray}

In order to show that also $\langle \Omega|\int d^4 x (T^{\mu}_{\nu})_1|\Omega\rangle=0$ we simply note that one can write:
\begin{eqnarray}
\prod_n\frac{1}{-p_n^2+m^2}=\prod_n\frac{1}{-p_n^2}\prod_n\frac{1}{1-\frac{m^2}{p^2}}.
\label{re776640}
\end{eqnarray}
Then using,
\begin{eqnarray}
(\det[1-\frac{m^2}{p^2}])^{-1}=\exp[-{\rm Tr}\ln(1-\frac{m^2}{p^2})]
\label{res554664}
\end{eqnarray}
and noting that the trace actually means integration over the momenta we conclude that this expression is totally independent of individual momenta. Since the first factor in the right hand side of Eq.
(\ref{re776640}) cancels the contribution of fermion momenta then the full partition function is independent of momenta and we can apply the findings of section II (which are based on similar reasoning but for
the general case).  Thus:
\begin{eqnarray}
\langle \Omega|\int d^4 x (T^{\mu}_{\nu})_1|\Omega\rangle=0
\label{tns5546}
\end{eqnarray}
Eqs. (\ref{proof65775}) and (\ref{tns5546}) show that for the simple supersymmetric example that we considered:
\begin{eqnarray}
\langle \Omega|\int d^4 x {\cal H }|\Omega\rangle=0.
\label{tns5546}
\end{eqnarray}

\end{appendix}

\section*{Acknowledgments} \vskip -.5cm

The work of R. J. was supported by a grant of the Ministry of National Education, CNCS-UEFISCDI, project number PN-II-ID-PCE-2012-4-0078.


\begin{thebibliography}{15}
\bibitem{Jaffe1} R.L. Jaffe, Phys. Rev. D {\bf 72}, 021301 (2005), hep-th/0503158.
\bibitem{Jaffe2} N. Graham, R. L. Jaffe, V. Khemani, M. Quandt, M. Scandurra, H, Weigel, Nucl Phys. {\bf B 645}, 49-84 (2002), hep-th/0207120.
\bibitem{Bass} S. D. Bass, Mod. Phys. Lett. A {\bf 30} 22, 1540033 (2015), arXiv:503.05483.
\bibitem{Sola} J. Sola, AIP Conf. Proc. 1606, 19-37 (2014), arXiv:1402.7049; J. Phys. Conf. Ser. 453, 012015 (2013), arXiv:1306.1527.
\bibitem{Carroll} S. M. Carroll, Living. Rev. Rel. 4, 1 (2001), astro-ph/0004075.
\bibitem{Bousso} R. Bousso,  Gen. Rel. Grav.  {\bf 40}, 607-637 (2008), arXiv: 0708.4231,
\bibitem{Cohn} J. D. Cohn, Astrophys. Space. Sci. {\bf 259}, 213-234 (1998), astr-ph/9807128.
\bibitem{Sahni} V. Sahni, A. A. Starobinsky, Int. J. Mod. Phys. A, {\bf 9}, 373-443 (2000), astro-ph/9904398,
\bibitem{Weinberg} S. Weinberg, Rev. Mod. Phys.  {\bf 61}, 1-23 (1989), astro-ph/9610044.
\bibitem{Dyson} L. Dyson, M. Kleban, L. Susskind, JHEP {\bf 10}, 011 (2002), hep-th/020813.
\bibitem{Jora} R. Jora, arXiv: 1510.08218.
\bibitem{Riess} A. G. Riess eta al., Astron. J. {\bf 116}, 1009 (1998), atsro-ph/9805201.
\bibitem{Perlmutter} S. Perlmutter et al., Astrophys. J. {\bf 517}, 565 (1999), astro-ph/9812133.
\bibitem{Tegmark} M. Tegmark et al., Phys. Rev. D {\bf 74}, 123507 (2006), astro-ph/0608632.
\end{thebibliography}
\end{document}